\documentclass[journal=jpccck,articletitle=true,manuscript=article]{achemso}
\usepackage{bm}
\usepackage{epsfig}
\usepackage{graphicx}
\usepackage{amsmath}
\usepackage{amssymb}
\usepackage{gensymb}
\usepackage[usenames]{color}
\usepackage{color}
\usepackage{mathcomp}
\usepackage{epstopdf}
\usepackage{float}
\usepackage{caption}
\usepackage{subcaption}

\usepackage{xcolor}
\usepackage{ulem}
\newcommand\colorsout[1]{\bgroup \markoverwith{\textcolor{#1}{\rule[0.5ex]{2pt}{0.4pt}}}\ULon}

\title[]{\bf FeCoCp$_3$ Molecular Magnets as Spin Filters}

\author{P.N. Abufager}
\email{abufager@ifir-conicet.gov.ar}
\affiliation{ICN2 - Institut Catala de Nanociencia i Nanotecnologia, Campus UAB, E-08193 Bellaterra (Barcelona), Spain}
\affiliation{Instituto de F\'{\i}sica de Rosario, Consejo Nacional de 
Investigaciones Cient\'{\i}ficas y T\'ecnicas (CONICET) and Universidad Nacional 
de Rosario, Av. Pellegrini 250 (2000) Rosario, Argentina}

\author{R. Robles}
\affiliation{ICN2 - Institut Catala de Nanociencia i Nanotecnologia, Campus UAB, E-08193 Bellaterra (Barcelona), Spain}

\author{N. Lorente}
\email{nicolas_lorente001@ehu.eus}
\affiliation{ICN2 - Institut Catala de Nanociencia i Nanotecnologia, Campus UAB, E-08193 Bellaterra (Barcelona), Spain}
\affiliation{Centro de F{\'{\i}}sica de Materiales
CFM/MPC (CSIC-UPV/EHU), Paseo Manuel de Lardizabal 5, 20018 Donostia-San Sebasti\'an, Spain}
\affiliation{Donostia International Physics Center (DIPC), Paseo Manuel de Lardizabal 4, 20018 Donostia-San Sebasti\'an,
Spain}
\date{\today}

\begin{document}

\begin{tocentry}
%\begin{figure}[h]
\begin{center}
\includegraphics*[scale=0.35]{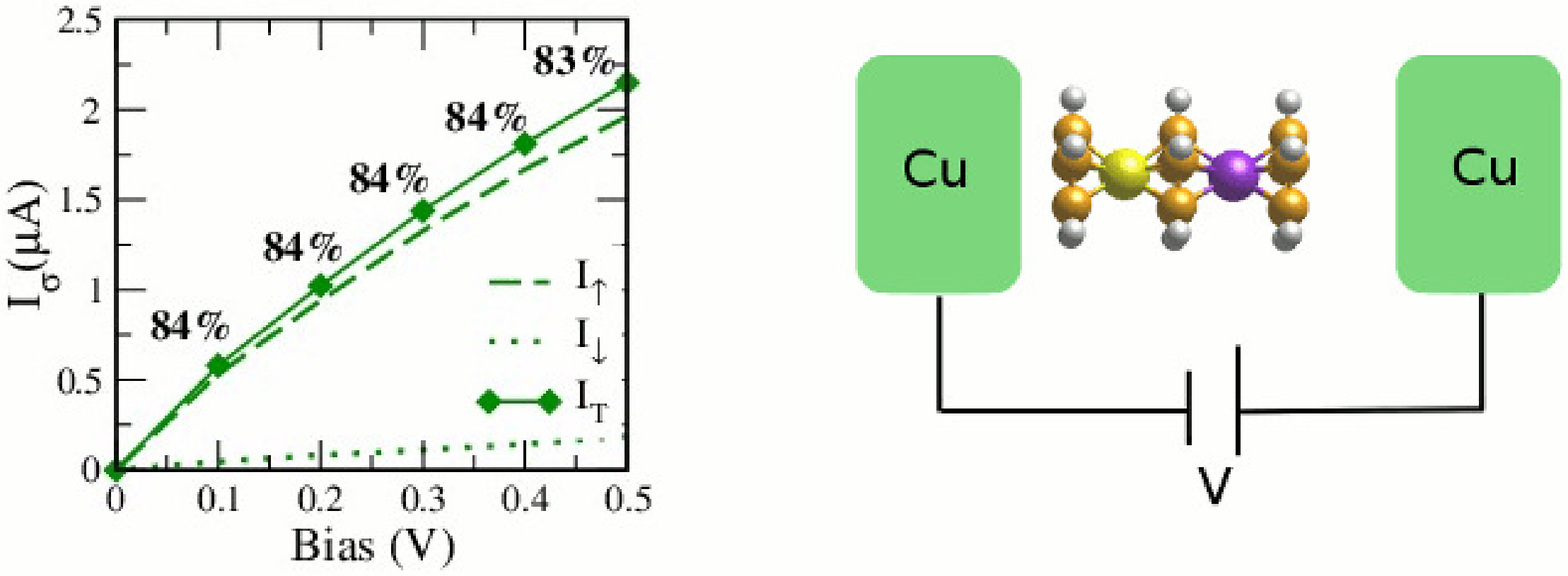}
\end{center}
%\caption{Table of Contents (TOC) Image} 
%\end{figure}
\end{tocentry}

\begin{abstract}
Metallorganic molecules have been
proposed as excellent spin filters in molecular spintronics because
of the large spin-polarization of their electronic structure. However,
most of the studies involving spin transport, have disregarded fundamental
aspects such as the magnetic anisotropy of the molecule and the excitation
of spin-flip processes during electron transport. Here,
we study a molecule containing a Co and an Fe atoms stacked between three
cyclopentadienyl rings that presents a large magnetic anisotropy and a S=1.
These figures are superior to other molecules with the same transition metal,
and improves 
the spin-filtering capacities of the molecule. 
Non-equilibrium
Green's functions calculations based on density functional theory
predict excellent spin-filtering properties both in tunnel and contact
transport regimes. However, exciting the first magnetic state drastically
reduces the current's spin polarization. Furthermore, a difference of
temperature between electrodes leads to strong thermoelectric effects
that also suppress spin polarization. Our study shows that in-principle
good molecular candidates for spintronics need to be confronted with
inelastic and thermoelectric effects.
\end{abstract}

\maketitle

\section{Introduction} \label{Introduction}

Molecular spintronics is a thriving field driven by advances in
shrinking electronic devices using molecules~\cite{Joachim} and by the
extraordinary properties of spin transport~\cite{Wolf,Wernsdorfer}. Not
only are molecules complex enough to attain dedicated functionalities,
but they are identically replicated and cheap to manufacture using
chemical synthesis. Recently, it has been possible to address
individual molecules while taking advantage of their hierarchical
growth to create structures of increasing complexity.~\cite{Barth}
Molecules can become fundamental pieces of the ever shrinking device
technology~\cite{RamanNature2013}. Additionally, molecules show
a great diversity of magnetic properties that can be successfully
tailored, such as spin-crossover molecules~\cite{SC}, molecular
magnets~\cite{Wernsdorfer}, spin-filtering molecules~\cite{HaoAPL2010},
molecular spin valves~\cite{sv} and molecular switches~\cite{HaoAPL2010}.
Molecular spintronics is then a rich field which promises scientific
and technological breakthroughs.

An interesting functionality that has been sought after in molecules
is the capability of selecting one spin to be transmitted in a
given spintronic device~\cite{Sen2010}. In order to achieve this,
the molecule presents spin-polarized frontier orbitals, with one of
the spins more coupled to the contacting electrodes. In this way,
the coupled molecular orbital has a larger contribution to electronic
transport, favoring the transmission of one spin species. Typically
the molecular spin polarization is achieved by using complexes where
the metallic atom (or atoms) present an open-shell configuration. The
ligand field of the rest of the molecule on the metallic atoms lead to
interesting physics: different spins can be present within the small
energy scale of the ligand field~\cite{SC}.  This is particularly true
in the case of a sizable spin-orbit coupling (SOC) and a spin larger than
1/2, because the ligand field creates a magnetic anisotropy due
to the SOC that can
fix the orientation of the molecular spin leading to the appearance
of molecular magnets.~\cite{Yamashita}  However, even in the absence
of a fixed magnetic-moment orientation, many different molecular
systems have been signaled as spin filters, because transport is
basically dominated by one of the electron's spins.  Indeed, recent
works~\cite{HerrmannJACS2010,HerrmannJCP2011} show that in the absence of
a magnetic center, radical molecules can be used leading to spin-polarized
electron transport.

Large spin polarizations have been predicted for the family of
molecules made from intercalated sequences of organic rings and
transition metals. Examples of these molecules are benzene-vanadium
ensembles~\cite{KoleiniPRL2007}, benzene-cobalt~\cite{Sen2010},
cobaltocene~\cite{liuJCP2007} and ferrocene and 1-D ferrocene-based
wires~\cite{ZhouJACS2008}.  Complete studies of different stacking of
cyclopentadienyl (Cp) and transition metals (TM) or benzene and transition
metals have also been performed~\cite{WangNL2008,BagretsJCTC2013}.
Stacking two different TM's has been less common. Some calculations suggest
that infinite sequences of stacked TM-Cp present exotic electronic
structure with different magnetic ordering depending on the used TM
atom~\cite{TMCp}.  Here, we propose a new molecular spin filter by
stacking an iron and a cobalt atom between three cyclopentadienyls
(Cp-Fe-Cp-Co-Cp).% based on recent reports that these molecules may be
%assembled using atomic manipulation techniques with a scanning tunneling
%microscope (STM).~\bibnote{Laurent Limot, \textit{Private Communication}.}
We performed non-equilibrium Green's functions (NEGF) simulations to
evaluate the transport properties of this molecule, CoFeCp$_3$,  based
on density functional theory (DFT).  As expected, the spin polarization
obtained in transport approaches 100\%. Moreover, the hybrid magnetic
structure of this molecule leads to a ferromagnetic coupling between
the magnetic centers, where most of the magnetization is localized
on the cobalt atom. Due to the
sizable spin-orbit coupling of Co, the Cp ring induce a sizable magnetic
anisotropy energy (MAE) which is very interesting for spin-filtering
applications. However, this same energy scale sets the 
energy scale for the first spin excitations
that can drastically reduce the spin-polarization in the
electron current~\cite{Gauyacq2012,Kepenekian2014}.
We evaluate here the effect of bias in reducing the spin polarization
as spin-flip processes become energetically accessible.

Our calculations show that transport takes place through the molecular
electronic structure based on its $\pi$-orbitals.  The broken-symmetry
electronic structure of CoFeCp$_3$ leads to frontier orbitals of
different nature and spin. In contact with metallic electrodes, only the
tails of the resonances caused by the molecule-electrode interaction
contribute to transport.  Hence, transmission changes rapidly with
energy near the Fermi energy which should lead to large thermoelectric
effects~\cite{CutlerPR1969,PaulssonPRB2003,ReddyScience2007}.
Moreover, the thermoelectric properties should be different per
spin, which can lead to spin currents even in the absence of charge
currents~\cite{Cornaglia2012,LuoScientificReports2013,Yang2014}.

\section{Theoretical methods}

In order to perform the calculations of this work, we have mainly
used two density-functional theory (DFT) packages. 
VASP \cite{Kresse1993a,Kresse1993b,Kresse1996a,Kresse1996b,Kresse1999,Hafner2008}
has been used to explore the adsorption of the CoFeCp$_3$ molecule
on the Cu(111) surface {and also its magnetic anisotropy}.
Geometrical effects when a second
electrode (another Cu(111) surface) was approached, have been evaluated
with VASP.  
However, the bulk of the calculations has been performed using
the {\sc Siesta} package~\cite{Soler}. These calculations confirmed the 
results obtained from VASP and permitted us to perform electronic
transport calculations using {\sc TranSiesta}.~\cite{Brandbyge2002}

We optimized the structure of the CoFeCp$_3$|Cu(111) interface, using
density functional theory (DFT) at the spin-polarized generalized
gradient approximation (GGA-PBE) level, as implemented in VASP
\cite{Kresse1993a,Kresse1993b,Kresse1996a,Kresse1996b,Kresse1999,Hafner2008}.
 In order to introduce long-range dispersion corrections, we employed
the so called DFT-D2 approach proposed by Grimme \cite{Grimme2006}.
We  used a plane wave basis set and the projected augmented wave (PAW)
method with an energy cut-off of 400~eV.  A 19-\AA~thick vacuum region
was used to decouple the surfaces of consecutive slabs in the supercell
approach used in VASP.  The surfaces were modeled using a slab geometry
with five Cu layers and a $3 \times 2\sqrt{3}$ unit cell. Such an unusual
unit cell have been chosen based on experimental data  for  ferrocene
(FeCp$_2$) molecules, which can be seen as one of the building blocks
for CoFeCp$_3$. Self-assembled monolayers of ferrocene shows a  $6 \times
2\sqrt{3}$ periodicity with two molecules per unit cell \cite{Limot2011}.
Published calculations yield that these two molecules do not interact
between them~\cite{Maider2015}.  Therefore, we decided to carry out our calculations using
a smaller $3 \times 2\sqrt{3}$, which is still large enough to prevent interactions
among adsorbed molecules.

During the geometry optimizations, we allowed for the relaxation
of all atoms of the molecule and of the two-topmost layers of the
Cu surface until the atomic forces were smaller than 0.02~eV/\AA.
A 7$\times$7$\times$1 k-point sampling of the first Brillouin zone was
performed using the Monkhorst-Pack method~\cite{Monkhorst1976}.

Transport calculations were carried out from first-principles with a
method based on nonequilibrium Green's functions (NEGF) combined with
DFT as implemented in the {\sc TranSiesta} package~\cite{Brandbyge2002}.
The open-boundary system is divided in three distinct regions breaking 
the periodicity along the transport direction. 
The central part is the scattering region and 
the other two regions are the semi-infinite left and right electrodes, 
formed by periodically repeating six layers of bulk copper.

The most favorable configuration after geometrical optimization of the
CoFeCp$_3$|Cu(111) interface was used to build the scattering region.
As illustrated in Fig.~\ref{scattering},
the scattering region was composed of one CoFeCp$_3$ molecule connected to
two Cu(111) surfaces, left and right, each formed by 8 active layers of
a $3 \times 2\sqrt{3}$ cell.  
It is important to stress that two-probe system geometries were obtained after
geometry optimizations using VASP. Dispersion corrections were described through the
semi-empirical DFT+D2 scheme.  Hence, the role of dispersion forces on the transport
results is implicitly considered trough the optimization of the junction's geometry.

For transport calculations, the valence electrons wave functions were
expanded in a basis set of local orbitals. A double-$\zeta$ plus
polarization (DZP) basis set was used to describe the
molecular states and
and a single-$\zeta$ plus polarization orbitals (SZP) basis set for
the copper electrodes.  Diffuse functions were also included
to describe surface electrons. The use of a DZP basis
set  to describe the molecular states
is mandatory in order to yield correct transmission
functions. Indeed, 
a SZP basis set  led
to a shift of  the main molecular peaks of $\sim$ 0.3 eV with respect to 
the DZP molecular peaks.
However, using a DZP for the full system does not alter
the transmission functions noticeably.
 Therefore, the
chosen basis set seems to be a good compromise between computational cost
and quality.  We employed the GGA/PBE functional~\cite{Perdew1996} and
norm-conserving Troullier-Martins pseudopotentials~\cite{Troullier1991}.
A 11$\times$11 in-plane k-point mesh was adequate to obtain sufficiently
accurate transport results.

The spin-polarized electron current $I_\sigma$ ($\sigma=\uparrow,\downarrow$, denoting majority a minority spin channels respectively)
 was calculated using the Landauer-Buttiker expression
\cite{Datta1995}: 

\begin{equation}
 I_\sigma=\frac{e}{h} \int_{-\infty}^{\infty} \tau_\sigma (\epsilon,V) 
\left[ f(\epsilon,\mu_L,T_L)-f(\epsilon,\mu_R,T_R) \right] d\epsilon . 
\label{corriente}
\end{equation}

\noindent where $\tau_\sigma(\epsilon,V)$ is the transmission function
for an electron of energy $\epsilon$ and spin $\sigma$ when the bias
voltage between the two  electrodes is V.  In eq.\ref{corriente}, $
f(\epsilon,\mu_\nu,T_\nu) =(1+\exp(\epsilon-\mu_\nu)/k_BT_\nu)^{-1}$
is the Fermi Dirac distribution of electrode $\nu$ ($\nu=L,R$, left and
right electrodes respectively) with temperature $T_\nu$ and chemical
potential $\mu_\nu$ (note that $V=(\mu_L-\mu_R)/e$).  The electron charge
is given by $e$ and Planck's constant by $h$.

In the linear-response regime, $I_\sigma$ can be approximated as \cite{Cornaglia2012}
\begin{equation}
I_\sigma \sim G_\sigma V + G_\sigma S_\sigma (T_L-T_R) 
\label{corriente.estimada}
\end{equation}
\noindent  where $G_\sigma$ and $S_\sigma$ are the spin-dependent conductance and Seebeck coefficient which are calculated at zero bias voltage ($V=0$) as 
\begin{equation}
G_\sigma=\frac{e^2}{h} K_{0\sigma}(E_F,T),
\label{seebeck1}
\end{equation} 
and
\begin{equation}
S_\sigma=-\frac{1}{|e|T} \frac{K_{1\sigma}(E_F,T)}{K_{0\sigma}(E_F,T)},
\label{seebeck2}
\end{equation} 
\noindent where $E_F=\mu_L=\mu_R$ is the Fermi level
and 
\[K_{\alpha,\sigma}(E_F,T)= -\int \! \frac{\partial
f(\epsilon,E_F,T)}{\partial \epsilon } \, (\epsilon-E_F)^\alpha  \,\,\,
\tau_\sigma (\epsilon,0) \,\,\, \mathrm{d}\epsilon\] 
with $\alpha=0,1$. The
total electronic conductance is given by $G=G_\uparrow+G_\downarrow$.

Finally, the  spin-filtering capabilities  of the molecular junction is
analyzed in terms of the spin polarization of the current, CP, defined as
\begin{equation} CP=(I_\uparrow-I_\downarrow)/(I_\uparrow+I_\downarrow)
\times 100.  \label{CP} \end{equation} When both the temperature
difference and bias voltage between left and right electrode are
zero (i.e. $V=0$ and $T_L-T_R=0$) the spin-filtering capacities
are evaluated using the spin polarization of the transmission
function at the Fermi energy. The corresponding quantity is called spin-filter
efficiency \cite{Wu2009,Maslyuka2010} and is defined as \begin{equation}
SFE=(\tau_\uparrow(E_{Fermi},0)-\tau_\downarrow(E_{Fermi},0))/(\tau_\uparrow(E_{Fermi},0)+\tau_\downarrow(E_{Fermi},0))
\times 100.  \label{SFE} \end{equation}

\section{Results and Discussions}

In this section, we analyze and discuss the results obtained for
CoFeCp$_3$ as a spin filter in the transport of electrons between
two copper electrodes. The section is divided in several subsections to
give a thorough view of the properties of this molecular device. The
first subsection analyzes the isolated molecule and compares it to
related molecules, explaining why CoFeCp$_3$ is a good candidate
for a spin-filter device. The second subsection analyzes the adsorbed
molecule on Cu(111). The third subsection is devoted to electron
transport in the elastic regime in the absence of thermoelectric effects,
both for tunneling and high-conductance regimes. The modification
of the spin-filtering capacities when spin-flip processes are
allowed is evaluated in the following subsection. This section is
finished by a detailed account of the effect of thermoelectric
effects in the properties of CoFeCp$_3$ as a spin filter. 

\subsection{Gas-Phase CoFeCp$_3$}
\label{gas}

\begin{figure}[t]
\begin{center}
 \begin{subfigure}{0.6\textwidth}
        \centering
        \includegraphics[width=0.4\textwidth]{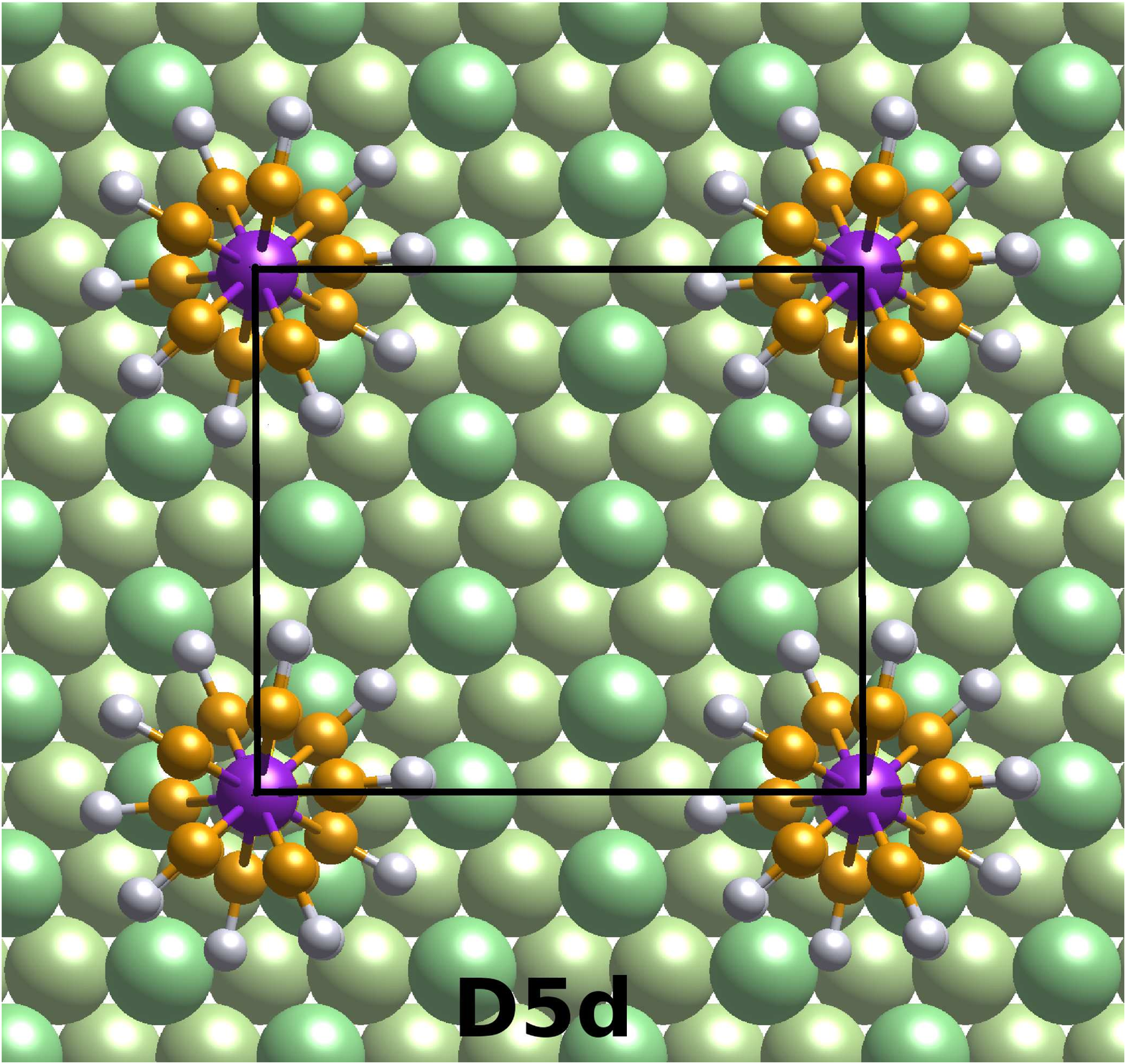}
        \includegraphics[width=0.4\textwidth]{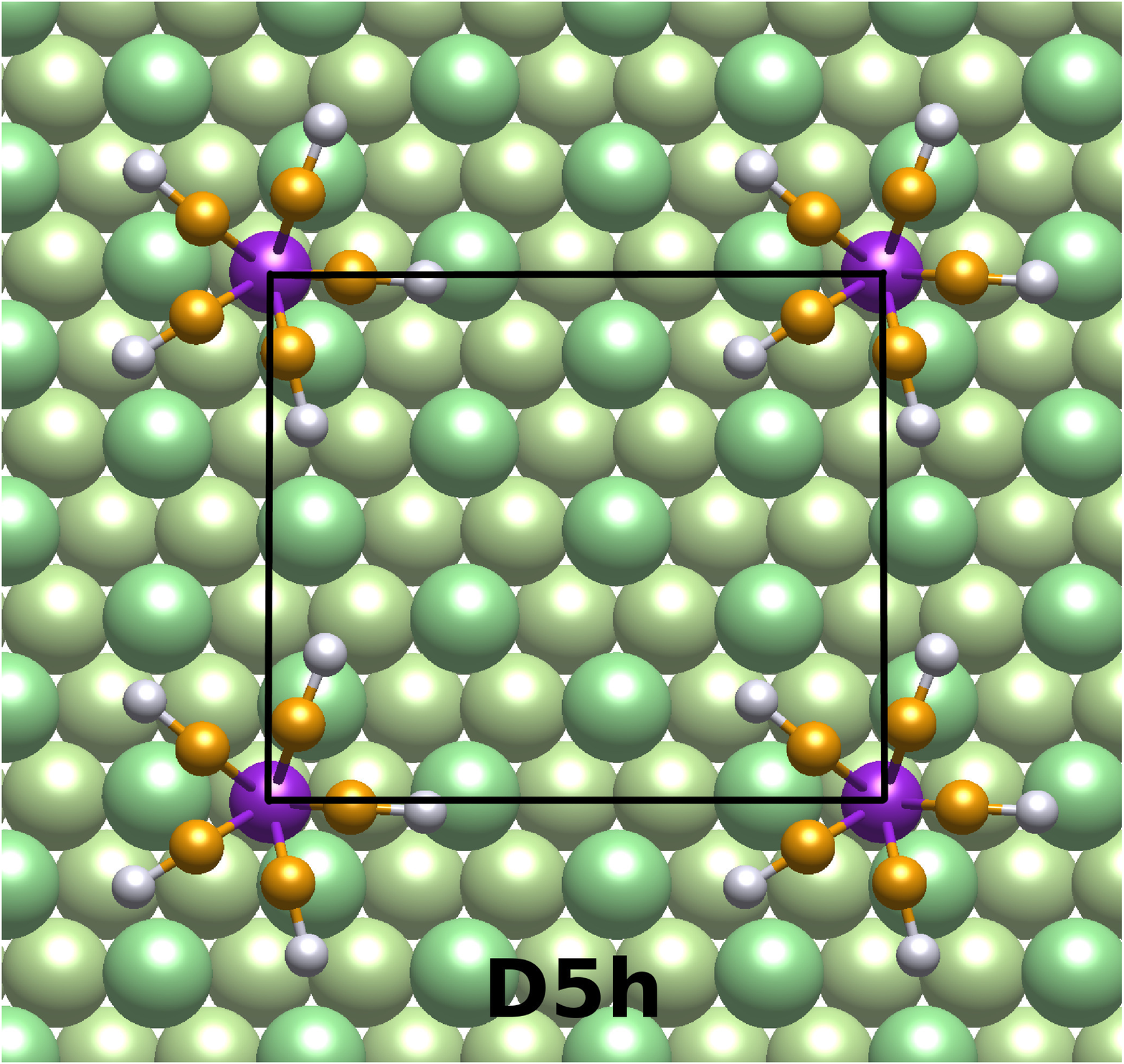}
        \captionsetup{justification=justified}
        \subcaption{}
    \end{subfigure}  %\hspace{1cm}
    \end{center}
\begin{subfigure}{0.6\textwidth}
        \centering
        \includegraphics[width=\textwidth]{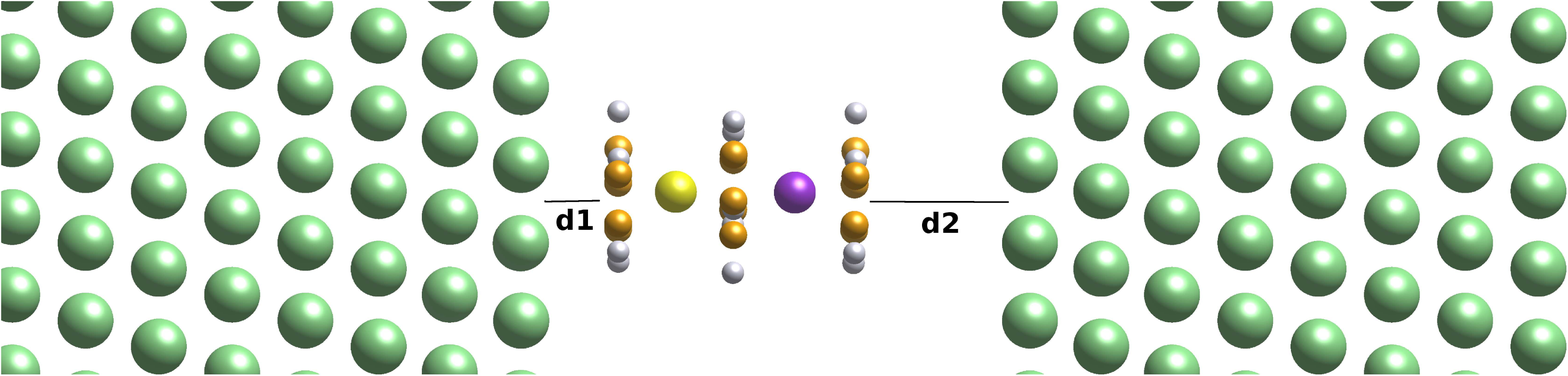}
        \captionsetup{justification=justified}
        \subcaption{}
    \end{subfigure}  
\caption{(a) Schematic top view of $D_{5d}$ and $D_{5h}$ conformers within the $3 \times 2\sqrt{3}$ unit cell. 
Yellow: Fe atom, violet: Co atom, orange: C atoms, grey: H atoms. 
The black lines represent the surface unit cell.
(b) Lateral view of the scattering region used in transport calculations.} 
\label{scattering}
\end{figure}

As shown in Fig.~\ref{scattering}, we considered two types of initial
structures for CoFeCp$_3$ molecules: eclipsed  and staggered
 (D$_{5h}$ and D$_{5d}$ symmetries, respectively).  In agreement with
previous results obtained for ferrocene, FeCp$_2$ \cite{Mohammadi2012},
the eclipsed  conformer is slightly more stable than the staggered one
(the computed energy difference is 58 meV).

\begin{figure}[t]
\begin{center}
\includegraphics*[scale=0.1]{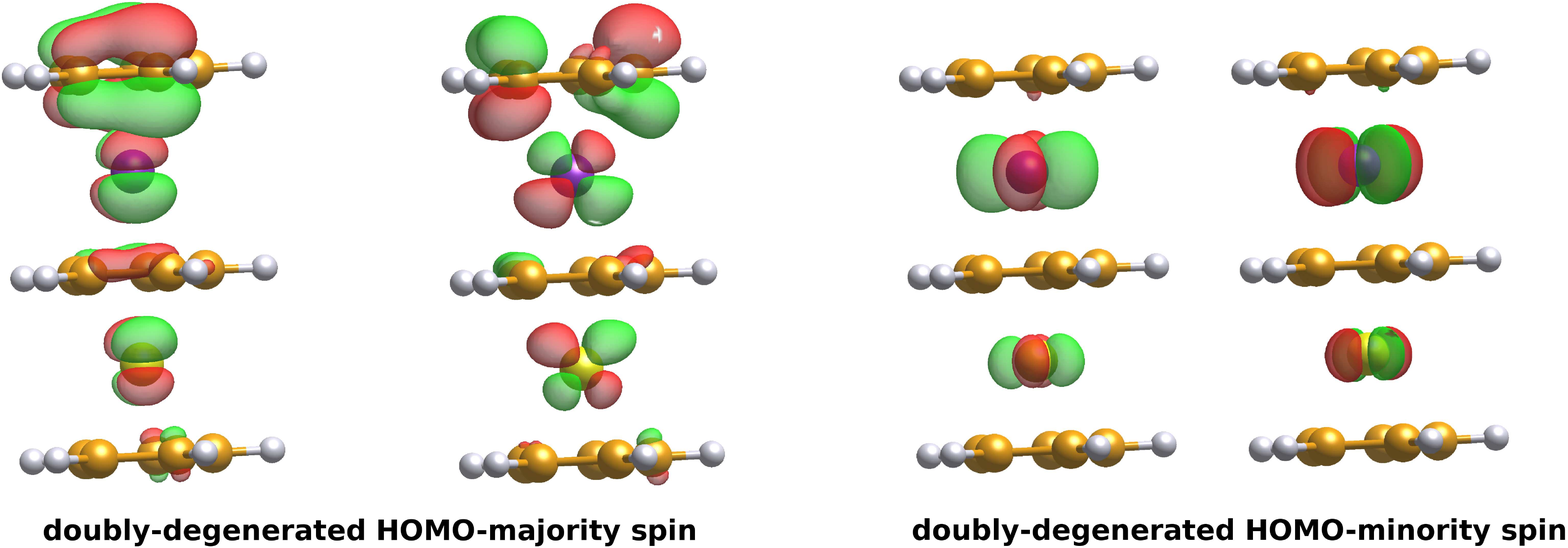} %\vline 
\vspace{2cm}
\includegraphics*[scale=0.1]{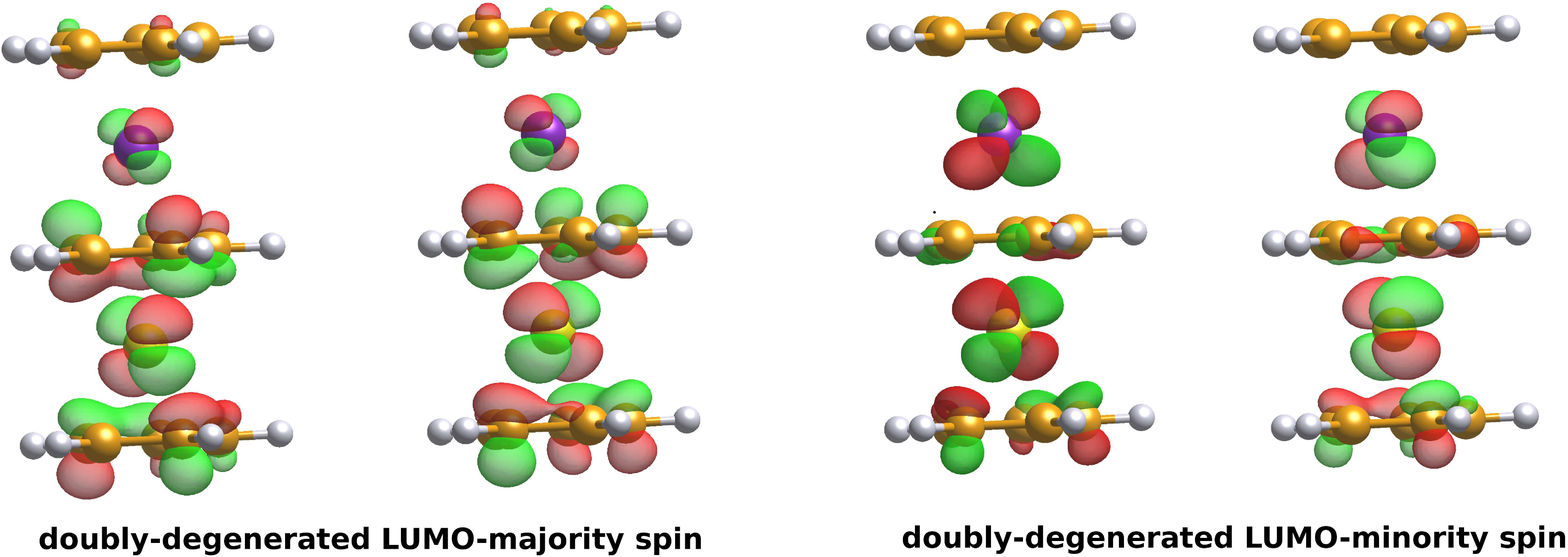} %\vline 
%\includegraphics*[scale=0.1]{Figures/LUMODOWN.eps} \\
%\includegraphics*[scale=0.2]{Figures/46down.eps} \\ 
%\subcaption{} 
\end{center}
\caption{Doubly-degenerated frontier molecular orbitals for the $D_{5h}$ conformer. Plotted isovalues are 10 $\%$ of the maximum ones. 
Red (green) indicates positive (negative) values of the real part of the wavefunction.} 
\label{orbitals}
\end{figure}

In both conformers, the ligand field splits the degenerated Co/Fe (TM)
$d$ levels into one $d_{z^2 }$ ($a_1$)  and two doubly-degenerated
$d_{xy}=d_{x^2-y^2}$ ($e_2$)  and $d_{zx}=d_{yz}$ ($e_1$) orbitals.
Depending on their symmetry and energy position, these orbitals mix to a
different degree with 2p states of the C  atoms. For instance, the highest
occupied  molecular orbitals (HOMO), for majority (HOMO$\uparrow$)
and minority (HOMO$\downarrow$) spin channels, schematically shown
in Fig.~\ref{orbitals}, have $\sim$ 50\% TM-$e_1$ and $\sim$ 90\%
TM-$e_2$ character, respectively.  On the other hand,  lowest unoccupied
orbitals (LUMO) for majority (LUMO$\uparrow$) and minority spin channels
(LUMO$\downarrow$) (see Fig.~\ref{orbitals}) present $\sim$ 50\% and 75
\% TM-$e_1$  character, respectively.  This picture agrees well with the
ligand-field splitting of the $d$-electron manifold in $D_5$-symmetry.

The Cp ligands roughly contain one electron. Hence, the TM atoms
approximately are in $d^6$ (Fe) and $d^7$ (Co) configurations, see
Table~\ref{table1}.  The lowest-energy conformation corresponds to the
low-spin one, hence filling the ligand-splitted $d$ levels for Fe and
Co leads to a spin 1 molecule. This is confirmed by our calculations,
regardless of the used exchange-and-correlation functional. From
this picture, we see that Co will host the spin one, and Fe will have spin
zero. This is in agreement with the zero spin of ferrocene. However,
cobaltocene (CoCp$_2$) is spin 1/2. The difference stems from the
presence of a Cp between Fe and Co in CoFeCp$_3$. Indeed, CoFeCp$_3$
is not a ferrocene plus a cobaltocene.  Plotting the spin distribution
for CoFeCp$_3$, we confirm the above results: spin is largely localized
on the Co atom, and the Fe atom is basically not magnetic.

The large spin-orbit coupling of Co, leads to a sizable MAE induced by
the Cp's ligand field. We have evaluated the MAE and we obtain that the
Co-Fe axis is a hard axis. This means that the magnetic moment of the
molecule lies in a plane parallel to the Cp's. The transversal anisotropy
is negligible. Hence the magnetic moment is not fixed in a particular
direction in the Cp's plane. The MAE is 1.64 meV for both
conformers. This is the energy needed to change the magnetic moment from the
easy plane to the hard axis. Since the magnetic moment corresponds to
$S=1$, the molecular ground state is doubly degenerate and formed by the
spin components $|S_z|=1$. The first excited state is $S_z=0$. Hence,
the magnetic moment will be localized in the Cp's plane as long as
the bias between electrodes is not large enough to flip the spin from
$|S_z|=1$ to $S_z=0$ as will be discussed below. 
 These
results have been obtained in the gas phase and are, in principle, not
valid for the adsorbed molecule. As we will see in the next section, the
molecule is basically physisorbed on Cu(111) without charge transfer or
any interaction from the substrate other than dispersion forces. Hence,
we expect that the gas-phase MAE be a good approximation to the MAE of
the spin-filter device.

These data indicate that CoFeCp$_3$ is a small molecule with an
important spin that is fixed to a plane contained by the Cp ligands,
with a pinning energy (MAE) of 1.64 meV.  Hence, the molecule can in
principle polarize an electronic current to a direction perpendicular
to the axis of the molecule. It is interesting to compare this molecule
with similar molecules. Co$_2$Cp$3$ or Fe$_2$Cp$_3$ will not be good
spin filters~\cite{Kepenekian2014}. The presence of an odd number of
Cp leads these molecules to present a low-spin configuration $S=1/2$,
which is not subjected to any magnetic anisotropy and cannot be
molecular magnets. Molecules with an odd number of Cp's and only
one type of TM atom such as Fe or Co, will probably not be good spin
filters either because they show antiferromagnetic coupling
with its corresponding $S=0$ ground state. Infinite chains of CoCp~\cite{TMCp} 
also show
antiferromagnetic ordering and hence a $S=0$ ground state. The case of
FeCp chains is more complex. For short molecules, the ground state is
the low-spin configuration $S=0$, however as the chain grows
larger, a half-metallic ferromagnet develops that can eventually be
an excellent spin filter~\cite{TMCp}. Here, we propose something simpler, just a
CoFeCp$_3$ molecule.

\begin{table}[h!]
\begin{center}
\begin{tabular}{|c|c|c|}\hline 
 Element      & Total Charge (Mulliken)  & Magnetization ($\mu_B$) (Mulliken)   \\\hline
 Fe           &  6.687 (d states= 6.191) &    0.513 (d states= 0.475)            \\\hline
 Co           &  7.881 (d states= 7.314) &    1.731 (d states= 1.684)            \\\hline
 C      &  62.772                  &    -0.262 (-0.228 Cp in between)                            \\\hline
 H            & 14.667                   &     0.018                              \\\hline
\end{tabular}
\caption{Charge distribution and magnetization for the isolated molecule (evaluated
with {\sc Siesta} and Mulliken-charge analysis). Total magnetization is 2$\mu_B$ ($S=1$).
\label{table1}}
\end{center}
\end{table}

\newpage

\subsection{Adsorption of  CoFeCp$_3$ on Cu(111)}

As a first step we carried out full geometry optimizations for a single
FeCoCp$_3$ molecule  with the Fe-Co axis  initially located on the
high-symmetry sites of Cu(111): top, bridge, hollow-hcp, and hollow-fcc.
The most (least) stable final configuration corresponds to the molecule
adsorbed on the hollow (top) site at an average distance of 2.65 \AA~
(2.78 \AA)  from the surface. However, the energy difference between
top and hollow adsorption sites is only 65 meV.  Since the computed
equilibrium points are spatially very close we do not expect to have
large energy barriers between the points, leading to an overall small
diffusion barrier.

Charge population calculations using the Bader scheme~\cite{Bader} point
to negligible charge transfer between molecule and surface. In addition,
neither the geometrical structure nor the electronic characteristic of
the molecule seem to be strongly affected by the adsorption process. As
a result, the adsorbed molecule maintains its gas-phase electronic and
magnetic properties.  This is further corroborated by a deep analysis
of the contributions to the total adsorption energy.

The main contribution to the adsorption energy, E$_{ads}$, comes
from dispersion forces (E$_{vdW}$).  Indeed, the evaluated adsorption
energy on the hollow site is E$_{ads}\sim -1.19$ eV. The contribution
to this adsorption energy is mainly due to the van der Waals component,
E$_{vdW} \sim-1.28$ eV that is reduced to the final E$_{ads}$ value by
the repulsion with the electronic cloud of the surface.  Interestingly,
if vdW interactions are turned off in the calculations, the molecule
feels the repulsive forces and reaches an adsorption distance of 3.22
\AA~with a very small adsorption energy ($E_{ads} \sim -0.13$ eV)
that is probably not meaningful. Nevertheless, these results show that
the molecule binds solely by the action of van der Waals forces.

\subsection{Transport properties of CoFeCp$_3$ on Cu(111)}

In the present section, we show the results of our electron transport
calculations for a CoFeCp$_3$ molecule between two Cu(111) electrodes
with special emphasis on spin filtering. The first results correspond to the
electron transmission across the molecular junction at zero bias. First,
the tunneling regime is analyzed, where the right electrode is kept at
a distance much larger than the adsorption one. Then, we analyze the
contact regime, also at zero bias. The third subsection explores bias
effect in the more interesting case of the contacted junction. And
finally, motivated by the slopes of the transmission function at the
Fermi energy, we compute the behavior of the molecular junction with
respect to a temperature gradient and the related thermoelectric effects.

All the results of this section have been evaluated for the eclipsed
(D$_{5h}$) molecular conformer.  The very similar data about the staggered conformer
can be found in the Supporting Information.

\subsubsection{Transport in the tunneling regime}

\begin{figure}[t]
\begin{center}
   \begin{subfigure}{0.95\textwidth}
        \centering
        \includegraphics[width=\textwidth]{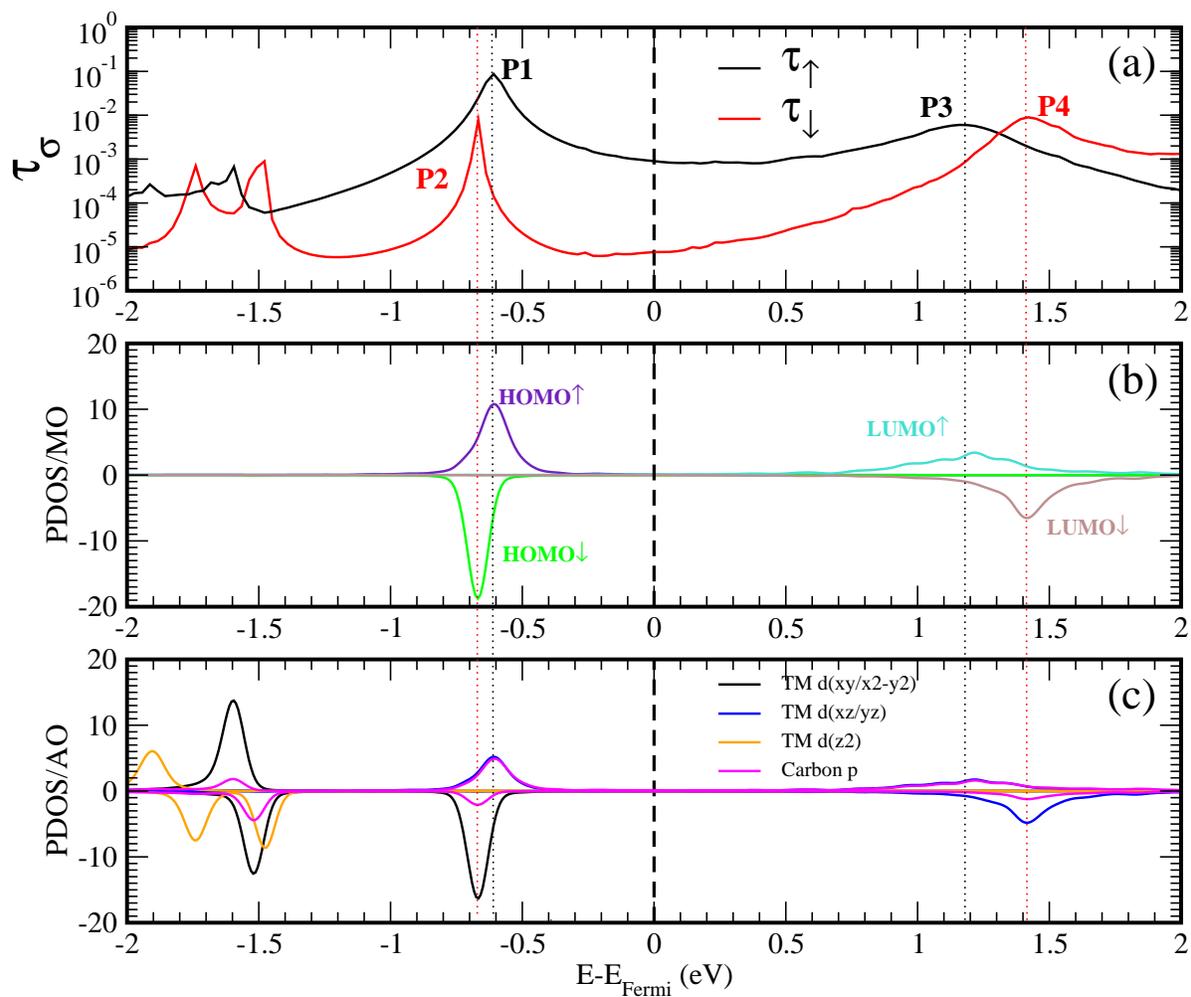}
        \captionsetup{justification=justified}
        %\subcaption{}
    \end{subfigure} \\ %vspace{15cm}
\end{center}
\caption{(a) Electron transmissions from the left to the right electrode as a function of electron energy referred to the Fermi energy. 
(b)Computed spin-polarized PDOS onto frontier molecular orbitals. (c) Computed spin-polarized PDOS projected onto TM-d and Carbon-p atomic orbitals.
Tunneling regime d$_2$=5.15 \AA. Total magnetization$=$ 2 $ \mu_B$. 
Mulliken charge transfer to the molecule = 0.098 $e$} 
\label{d2_5_norot}
\end{figure}

Figure~\ref{d2_5_norot} (a) shows the transmission spectra at zero
bias for a left-(right-)electrode-molecule distance of $d_1=$2.65
\AA~($d_2=$5.15 \AA). 
We approximate the zero-bias conductance by
the transmission at the Fermi level, $E_F$. Hence, the conductance
is $G(E_F)=4.53 \, \times \, 10^{-3} G_0$ where $G_0=2 e^2/h$ is the
quantum of conductance.  The small value of $G(E_F)$ shows that this
setup corresponds to the tunneling regime.

Concerning the effect of the electrodes on the geometry and electronic
properties of the molecule, we conclude that it seems to be very small. 
Within this geometry i) the structural parameters of the molecule are very similar
to the ones obtained in gas phase   ii) the total charge transfer to the molecule
is very modest (0.098 e) and iii) the total magnetization of the molecule is 2 $\mu_B$
where the partial contributions coming from Fe, Co, C and H 
(i.e 0.684 $\mu_B$, 1.567 $\mu_B$, -0.271 $\mu_B$ and 0.022 $\mu_B$, respectively)
are close to the values reported in Table I for the isolated molecule 

Figure~\ref{d2_5_norot}  shows the extraordinary spin-polarization
induced by the molecule. The majority-spin channel transmission
($\tau_\uparrow(E_{Fermi})$) is roughly two orders of magnitude higher
than the minority-spin one ($\tau_\downarrow(E_{Fermi})$). As a result,
the $SFE$ given by eq.~\ref{SFE} approaches 100 $\%$ (more precisely,
$SFE = 98\%$).

The transmission of Fig.~\ref{d2_5_norot} (a) implies that transport
is mainly determined by the hybridization of surface electronic states
with the frontier molecular orbitals.  To get a deeper understanding
of the different features observed in the transmission function, we
plot the density of states projected (PDOS) onto the
frontier orbitals that we analyzed above, 
namely, the doubly-degenerated
HOMO's and LUMO's.
These PDOS are depicted in Fig.~\ref{d2_5_norot} (b). 
The PDOS peaks perfectly match the transmission ones, permitting us
to identify them. 
\bibnote{The clear correspondence between peaks (marked P1, P2, P3 and P4 in
Fig.~\ref{d2_5_norot} (a)) on the transmission function and the PDOS,
Fig.~\ref{d2_5_norot}~(b-c), makes it possible to assign P1 at
-0.61 eV (P2 at -0.67 eV) and P3 at 1.18 eV (P4 at -1.42 eV) peaks to the
transmission trough HOMO$\uparrow$ (HOMO$\downarrow$) and  LUMO$\uparrow$
(LUMO$\downarrow$), respectively. The Fermi  level is
found to lie closer to HOMO states than LUMO states for both spin
channels. However, the linear-response regime is controlled
by the LUMO of both spins. We expect then
that a low-bias scanning tunneling microscope image will
be an image of the molecular LUMO.}  
Moreover, we can explain the spin-polarization
as due to the different spatial extend of the molecular orbitals
in each of the spin channels and the corresponding overlap
with the electrodes. Hence, the spin-polarization is rather an effect
of the geometry of the molecular orbitals at play rather than 
due to a spin-polarized density of states.

Projecting the density of states onto atomic orbitals is also instructive.
Figure~\ref{d2_5_norot} (c) depicts  the PDOS onto the atomic
TM-d and Carbon-p states. This permits us to corroborate the above conclusion.
Indeed, we can see that while the HOMO$\uparrow$ has a large component
on Carbon-p states, the HOMO$\downarrow$ is basically a TM-d orbital. This
same conclusion, but for different orbitals, is deduced from the LUMO
composition~\bibnote{The narrower-width and lower-height peak found for  P2
as compared to the  P1 one  can
be understood by  close inspection of the involved molecular orbitals. As
it was previously discussed, HOMO$\downarrow$ (P2) is basically an $e_2$ orbital 
($d_{xy}-d_{x^2-y^2}$ perpendicular to the transport direction) orbital with its
electronic cloud completely localized on the TM-centers, this leads to a
poor hybridization with the electrodes and low transmission probability.
On the other hand, the HOMO$\uparrow$ (P1) is more delocalized and it is built
from $e_1$ ($d_{xz}, d_{yz}$)  TM-d orbitals. As a result
its hybridization with the electrodes as well as its transmission
probability are higher.
Concerning peaks P3 and P4, it is interesting to note that although
both show similar transmission probabilities, the P3 structure
has a much broader energy range than the P4
one. Such features can be understood taking into account that  although,
LUMO$\uparrow$ (P3) and LUMO$\downarrow$ (P4) are both built from TM-d($e_1$) orbitals,
LUMO$\downarrow$ is much more localized about the TM-centers.
As a result, its hybridization with the surface electrodes is lower than for the
LUMO$\uparrow$.}. We can then conclude that the larger
contribution to the electronic current of the majority
spin ($\uparrow$) orbitals is due to the contribution of Carbon-p states,
and hence of the $\pi$-orbitals of the Cp ligands revealed in Fig.~\ref{orbitals}.

\begin{figure}[t]
\begin{center}
   \begin{subfigure}{0.95\textwidth}
        \centering
        \includegraphics[width=\textwidth]{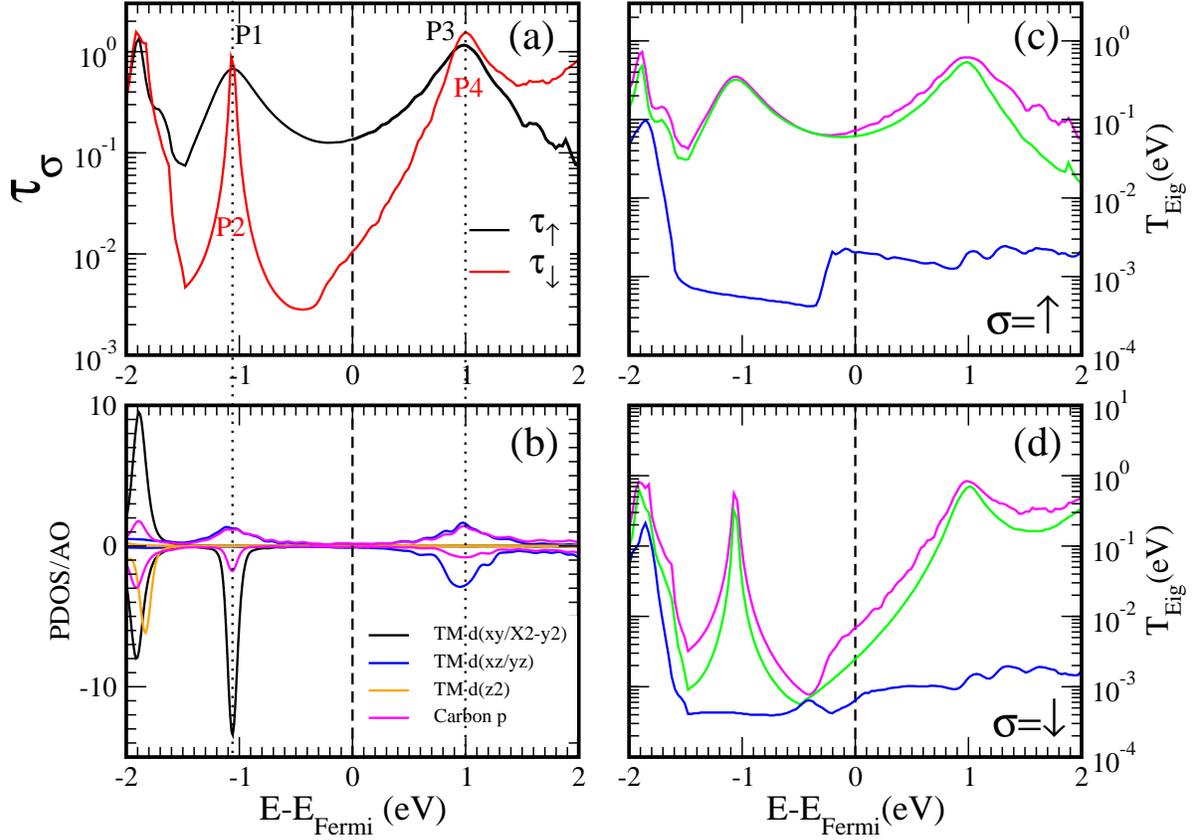}
        \captionsetup{justification=justified}
        %\subcaption{}
    \end{subfigure} \\ %vspace{15cm}
\end{center}
\caption{Contact regime d$_2 \sim$ 2.57\AA. (a) Electron transmissions from the left to the right electrode as a function of 
electron energy referred to the Fermi energy. (b) the PDOS onto TM-d  and Carbon-p states.
(c-d) Eigenchannels contribution to the average transmission for majority and minority spin channels
(the color-code only denotes the ordering of the transmission eigenvalue at each electron energy and
may not be connected to the symmetry or other common properties of
the eigenchannels).Total magnetization$=$1.92$\mu_B$.
Mulliken charge transfer to the molecule = 0.177~$e$.} 
\label{d2_2.5_norot}
\end{figure}

\subsubsection{Transport in the contact regime}

To mimic the contact regime, we approach the right electrode to the
molecule at a distance of $d_2=$2.57 \AA  ($d_1=$2.72 \AA) \bibnote{This
configuration is 0.01 eV more stable than the one with $d_1=d_2=2.65$~\AA~where 
2.65~\AA~is the average distance of the molecule on the surface}.
The present electrode configuration has been chosen so as not to exert
any pressure on the molecule. Hence, the electrodes induce negligible distortions
of the molecular geometrical parameters.  The charge transfer to the molecule is
still small (0.177 e)  and magnetic moment of the molecule reaches 1.895  $\mu_B$ 
(with 0.478 $\mu_B$, 1.630 $\mu_B$,  -0.234 $\mu_B$ and 0.021 $\mu_B$ 
for Fe, Co, C and H, respectively, 
which are similar to the ones described in Table 1).  
Overall,  this analysis and the one described for tunneling conditions  
indicate a modest effect of the electrodes and vdW-forces on the properties of
the molecule for the two-probe system.

Figure~\ref{d2_2.5_norot} (a) shows the transmission 
at zero bias as a function of the electron energy. 
As expected the transmission is larger than the transmission in the tunneling regime, 
leading to a total conductance at the Fermi level of
$G(E_F)=0.073 G_0$.  At the Fermi level,  the majority spin channel
exhibits a transmission probability one order of magnitude higher than the minority
spin channel.  Thus, the molecule maintains its spin-filter character ($SFE =
86 \%$).

For both spin channels, eigenchannel analysis~\cite{eigenchannel}
shows that two scattering states provide the major contribution to
the transmission function in the whole energy range. In particular,
the two most contributing scattering states provide very similar
contributions at the energies corresponding to peaks  P1, P2, P3 and
P4 (see Fig.~\ref{d2_2.5_norot}(b,d)).  Although, it is  difficult to
identify eigenchannels by visualizing them \bibnote{Due to the fact
that the eigenvalues of the transmission function corresponding to
eigenchannels at k$_{xy}=(0,0)$ do not provide the major contribution
to the total transmission}, the perfect energy alignment between these
peaks and the ones observed in the PDOS (Fig.~\ref{d2_2.5_norot}~(c))
allows again to assign   P1 (P2) and P3 (P4) peaks to transmission trough
HOMO$\uparrow$ (HOMO$\downarrow$) and  LUMO$\uparrow$ (LUMO$\downarrow$)
molecular orbitals.  Moreover, assuming Breit-Wigner-like resonances
for the transmitting MO, we have fitted the corresponding transmissions
with Lorentzian functions~\cite{Datta1995} which
permits us to confirm that the peaks in the transmissions
nicely corresponds with the molecular levels in the PDOS
(see Supporting Information).  Moreover, the Lorentzian fitting, albeit
imperfect, shows that the LUMO transmission dominates at the Fermi energy
for both spin channels. As the electrode approaches the molecule, the
contributions of the LUMOs grow, with no reversal of molecular
character in the electron transmission.

As can be seen, the 4 frontier-orbital peaks shift to lower energies with
respect to their energy position  in the tunneling regime.  This is due
to the enhancement  of the molecule-electrode interactions which
also induce a more pronounced broadening  of the involved molecular
levels. Interestingly, the largest hybridization is observed for P1
where the broadening increases roughly a factor of 6. This
behavior can be traced back to the larger overlap
of the HOMO$\uparrow$ with the approaching electrode. 

Our transport calculations carried out for the staggered
conformer show that the symmetry of the molecule does not affect the
spin-filter character of the molecular junction (see Supplementary
Information for more details.)
 
\begin{figure}[t]
\begin{center}
   \begin{subfigure}{0.8\textwidth}
        \centering
        \includegraphics[width=\textwidth]{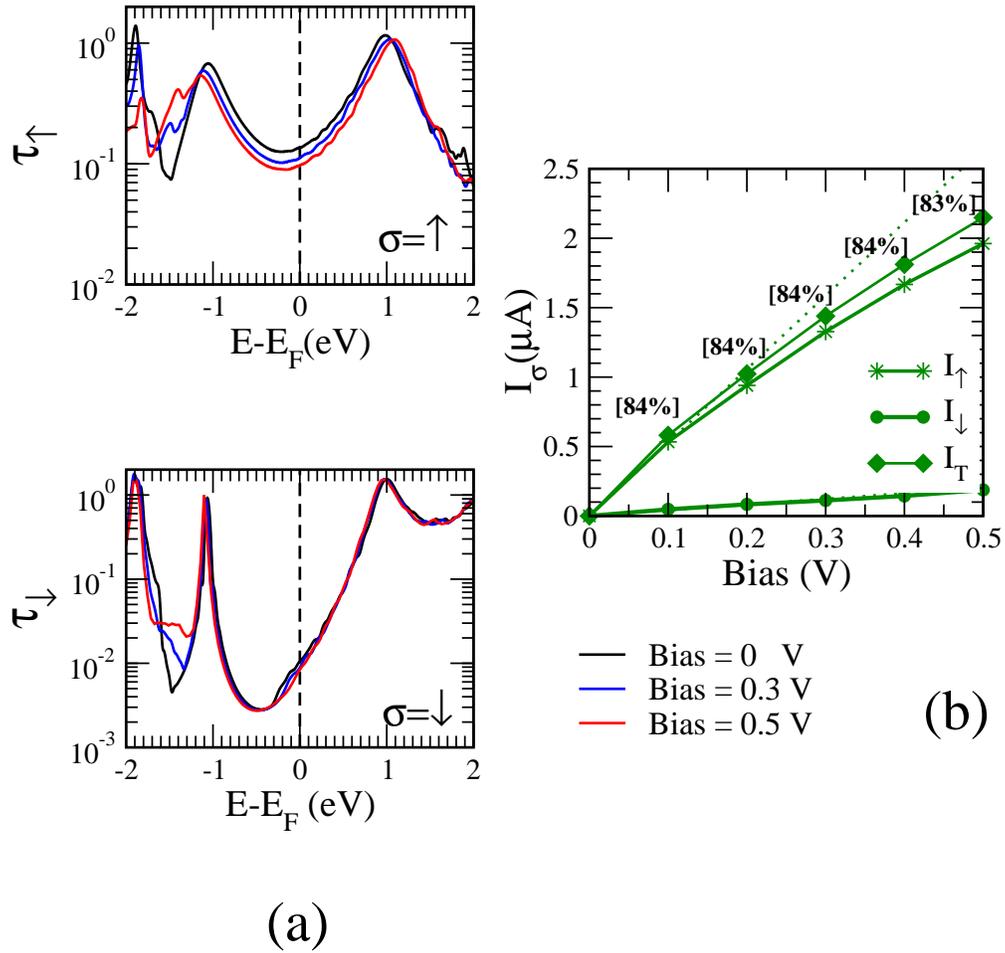}
        \captionsetup{justification=justified}
        %\subcaption{}
    \end{subfigure} \\ %vspace{15cm}
\end{center}
\caption{(a) Electron transmission as a function of electron
energy for 0, 0.3 and 0.5 V applied between the two metallic electrodes. The majority
spin (upper panel) shows that the HOMO-LUMO gap is broadened as
the bias is increased and there is a small drop of transmission at
the Fermi energy. For the minority spin (lower panel)  the shifts are
negligible and there is a very minor bias dependence. 
(b) I-V characteristics in nA and V. The SFE is shown in brackets. 
Dot lines indicates the results obtained with the linear approach $I_\sigma=G_\sigma \, V$ 
} 
\label{Current_Bias}
\end{figure}
 
\subsubsection{Finite-bias results}

The above results imply that FeCoCp$_3$ is a good
spin filter in the linear-response regime.
In this section, we go beyond the linear-response regime.
We computed the electron current  for both spin channels
(I$_\uparrow$,I$_\downarrow$) as a function of the applied bias using
eq.~\ref{corriente}.  

Figure~\ref{Current_Bias}(a) shows the electron transmission that enters
the Landauer equation, eq.~\ref{corriente}, evaluated for three different bias.
The upper panel shows the majority spin transmission. We find that the
HOMO-LUMO gap increases with bias and the transmission in between the 
two peaks decreases. For the minority spin (lower panel)
the bias effect is negligible. Overall, the effect of the bias is small
and using the zero-bias transmissions seems justified. Nevertheless,
it is interesting to both understand why the bias effect is small and
why the effect is not noticeable for the minority spin. 

The effect is small because the molecule is basically bound by
dispersion forces, hence the molecular electronic structure presents
small perturbations from the electrodes.  The presence of an external
electrical field acts on the polarization of the molecule. Here, the
fields are so small that this effect is negligible. The flowing of a
current through the molecule is a larger effect, leading to a change in
the steady-state charge of the molecule. However, the HOMO stabilizes by
trapping a very small amount of charge and in the same degree the LUMO
empties, contributing to an almost zero change in charge state. This
leads to a small opening of the HOMO-LUMO gap.

As we have previously seen, the minority-spin molecular
orbitals are less coupled to the substrate. Hence, the opening
of the HOMO-LUMO gaps is negligible. 
Interestingly, the effect of the bias on the magnetic moment of the molecule is negligible
(i.e it goes from 1.895 $\mu_B$ to 1.860 $\mu_B$ when the bias  increases 0.5 V,
see Tables 3-4 in the Supporting Information).

Figure~\ref{Current_Bias}(b) shows the current computed using the Landauer
equation, eq.~\ref{corriente}.  In the linear regime the majority-spin
current, I$_\uparrow$  is appreciably larger than I$_\downarrow$ due
to the higher conductivity for majority than for minority spin channels
(0.0675$G_0$ vs 0.0055$G_0$).  
Such a large difference between I$_\uparrow$
and  I$_\downarrow$ is still observed as the voltage further increases.
As a result, FeCoCp$_3$ acts as a spin filter in the whole bias voltage
range with a large current polarization, CP $\sim$ 84 $\%$.  The inclusion
of bias in our calculations does not change the conclusion that this
molecule is an excellent spin filter with a CP close to 90$\%$.

These results are in contrast with the ones of Ref.~\cite{Cakir2014}
where they find that the CP in a Fe-C$_{70}$C$_{70}$-Fe junction goes
from 78\% at zero bias to  20\% at 0.5V. The large difference between our
results and theirs can be traced back to the very different interaction
of the molecules with the electrodes. While in our case the molecule is
physisorbed by van der Waal forces, in their case, a strong covalent
interaction rules the charge flow through their Fe-C$_{70}$C$_{70}$-Fe
junction.

\subsection{Spin flip effects}
\label{flip}

The electronic current can yield energy to the molecular spin degrees of 
freedom, and hence change the spin state of the molecule. As a consequence
spin excitations can reduce the spin-filtering capabilities
of the device if the excited states corresponds to different
spin alignments. Let us briefly describe spin transport
through the FeCoCp$_3$ molecule.

The MAE of the molecule is 1.64 meV as described above. %in section~\ref{gas}. 
The molecular
axis aligning the Fe and Co atoms is a hard axis. Hence, the molecular
ground state corresponds to a spin of 1 in the easy plane described
by the Cp ligands which corresponds to a $S_z=0$ if
the molecular axis is taken as the $z$-axis. 
In this conditions, the electron spin is contained in the molecular
easy plane. As we have seen before, the spin-polarization with respect to an
axis on this plane will be very large, well above $80\%$ in all the cases analyzed above. 
Precession of the spin-polarization axis will be small, and the spin
current will be polarized in an arbitrary axis contained in the molecular plane.

A spin Hamiltonian can be written that reproduces the MAE for this $S=1$ molecule.
We can easily see that
\begin{equation}
\hat{H}_{spin}= D S_z^2.
\label{spinHamiltonian}
\end{equation}
In the present case the value of D is $1.64$ meV. From here we see that
the first excitation is indeed equal to 1.64 meV and it corresponds
to flipping the spin from $S_z=0$ to $|S_z|=1$, \textit{i.e.} from
the easy plane to the hard axis. Hence, electrons with energy above
the first-excitation threshold (biases above 1.64 mV) can flip the
molecular spin out of the easy plane if they flip their spin. A simple
calculation~\cite{Gauyacq2012} shows that the incoming electron has a
probability of 1/3 to flip its spin in the present case. As a consequence
the $CP$ goes from a value close to 100\% to 33\% when the absolute
value of the applied bias goes above 1.64 mV (in the case
where the intrinsic spin-polarization due to the electronic structure is
the $84\%$ of the previous section, the spin polarization above
the spin-flip threshold becomes $28\%$).

This description
is valid both for the tunneling and the contact transport regimes, since only the
molecular MAE and spin multiplicities enter it.

\subsubsection{Thermoelectric effects}

Motivated by the different ratios $\tau_\sigma^{'}/\tau_\sigma$
at the Fermi level for minority and majority spin channels,
we evaluate whether a spin-polarized thermopower current
can reduce the spin polarization of the total current.
This is of importance
because the spin filtering capacities may not be maintained 
in the presence of a temperature drop ($\Delta T= T_L-T_R$) 
across the junction. This physical situation can be reached
when the electrodes are contacted in a different way, and current
dissipation in the electrodes may lead to different temperatures.

\begin{figure}[t]
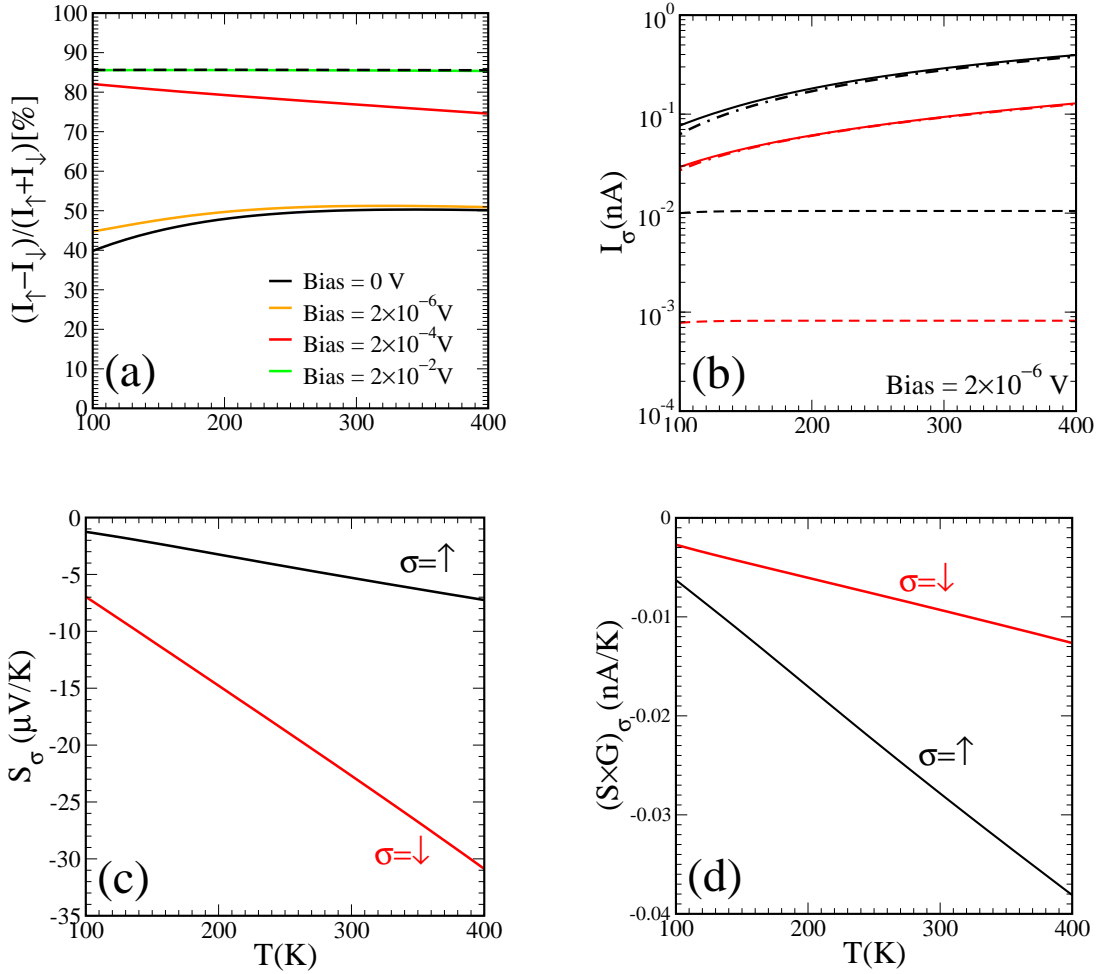

\begin{center}
   \begin{subfigure}{0.4\textwidth}
        \centering
        \includegraphics[width=\textwidth]{SFE_vs_T.eps}
        \captionsetup{justification=justified}
    \end{subfigure}  \hspace{1cm}
   \begin{subfigure}{0.4\textwidth}
        \centering
        \includegraphics[width=\textwidth]{Current_Voltage2d-6.eps}
        \captionsetup{justification=justified}
    \end{subfigure} \\ \vspace{0.5cm}
   \begin{subfigure}{0.4\textwidth}
        \centering
        \includegraphics[width=\textwidth]{S_vs_T.eps}
        \captionsetup{justification=justified}
    \end{subfigure} \hspace{1cm}
   \begin{subfigure}{.4\textwidth}
        \centering
        \includegraphics[width=\textwidth]{G_times_S_vs_T.eps}
        \captionsetup{justification=justified}
    \end{subfigure}  \hspace{1cm}
 \end{center}
\caption{
(a) Current polarization (CP) as a function of the average electrode temperature $T$ (i.e. $T=(T_L+T_R)/2$)
for different  Bias when $\Delta T= -10K$. The black dash line indicates the $CP$ values obtained 
when both electrode are at exactly the same temperature.
(b) Majority (black lines) and minority (red lines) spin polarized currents for 
an applied bias of $2\times10^{-6}$V ($\Delta T=-10 K$). 
Dash and dash-dot lines indicates $I_\sigma^{V}$ and $I_\sigma^{Th}$ contributions  
to the spin-polarized current $I_\sigma$ shown in solid lines; $\sigma=\uparrow,\downarrow$ (see text).
(c) Spin dependent Seebeck coefficient $S_\sigma$ ($\mu$V$/$K) as a function of the temperature (K).
(d) The Seebeck coefficient times the  linear conductance (nA/K) for the two spins as a function of the temperature (K).} 
\label{thermalcurrent}
\end{figure}

For this purpose, we take a temperature drop $\Delta T= -10K$ between
electrodes and compute the spin-polarized electron current, $I_\sigma$
with  $\sigma=\uparrow,\downarrow$ using eq.~\ref{corriente}, for
different Bias. The current polarization, $CP$ (eq.~\ref{CP}), obtained
in each case as a function of the average electrode temperature $T$
($T=(T_L+T_R)/2$) is shown in Fig.~\ref{thermalcurrent}(a). For the sake
of comparison, we plot the $CP$ values obtained when both electrodes are
at exactly the same temperature.

At zero ($Bias=0$) and extremely low bias  ($Bias=2\times10^{-6}$ V), we
see that thermal effects induce a drop of the CP value from $\sim$86$\%$
when both electrodes are at the same temperature to 40-50$\%$ in 
presence  of a small temperature gradient. However, the excellent
spin-filtering capabilities are restored as soon as the bias voltage
is slightly increase; the $CP$ reaches again 86$\%$ when the bias
is 0.02 V.

To understand such thermal effects on the current polarization, one simply
needs to make use of the linear-response limit of the spin polarized
electron current which tells us that $I_\sigma = I^V_\sigma + I^{th}_\sigma
= G_\sigma V + G_\sigma S_\sigma \Delta T$; $\sigma=\uparrow,\downarrow$
(see eq.~\ref{corriente.estimada}).
From this
expression, we can clearly establish two limiting behaviors: one dominated
by thermoelectric current $I^{th}_\sigma$ at low biases,  and  the
other one by the bias,  $I^V_\sigma$, when the bias
becomes larger than a critical bias, $V_c$, given by
$V_c \approx k_B \Delta T$. 

Let us focus on the first case, Fig.~\ref{thermalcurrent}(b), where $I_\sigma$
can be approximated by $ G_\sigma S_\sigma \Delta T$
($\sigma=\uparrow,\downarrow$).  Here, 
$CP$ is 
reduced to $(G_\uparrow S_\uparrow-G_\downarrow S_\downarrow)/(G_\uparrow
S_\uparrow + G_\downarrow S_\downarrow)$.  Hence, the Seebeck coefficient
($S_\sigma$) times the conductance ($G_\sigma$) for the two spin channels
are the key ingredients of the current polarization.  The
spin-dependent conductances   $G_\uparrow$ and $G_\downarrow$ with average
values 5250 \,\, nA$/$V  and  409 \,\, nA$/$V, respectively, barely change
in the studied temperature window.  In addition, the spin-dependent
Seebeck coefficient as a function of the electrode temperature plotted
in Fig. \ref{thermalcurrent}(c) shows that $|S_\uparrow|$ is roughly four
times lower than $|S_\downarrow|$. As a result, $|G_\uparrow S_\uparrow|$
is crudely three times larger than $|G_\downarrow S_\downarrow|$ (see
Fig. \ref{thermalcurrent}(d)) which explains the 40-50 $\%$  of current
spin polarization observed in Fig. \ref{thermalcurrent}(a).

With regards to the second case where $I_\sigma \sim G_\sigma V$,
the current spin polarization is here simplified to 
$CP \sim (G_\uparrow-G_\downarrow)/(G_\uparrow+G_\downarrow)$.
Therefore, the excellent spin-filtering capacities ($CP=86\%$) found 
in this case can be traced back to a much higher conductance 
for majority  than minority spin channels.

Summarizing, thermoelectric effects in this type of molecular junctions lead to a
strong suppression of the otherwise excellent spin-filtering properties
of the molecules when the electronic transport is governed by the 
thermoelectric current.

A different thermal effect is the one given by a homogeneous temperature. 
As the temperature rises, the direction of the molecular spin can change. 
Indeed, at $\sim$20 K, the ambient temperature is large enough to induce spin flips, 
similar to the spin-flips we have described in the previous section.

\section{Summary and Conclusions} \label{Conclusions}

Using DFT calculations together with a NEGF implementation of electronic
transport equations, we have evaluated the gas-phase, adsorption and
transport properties of a CoFeCp$_3$. The motivation to do so is the 
spin (S=1) of the gas-phase molecule, and its magnetic anisotropy
(MAE=1.64 meV).  These two properties are good characteristics for a
tentative molecular-based spin filter.

The molecular spin is largely localized on the Co atom,
and the Fe atom is basically not magnetic. This is due to the charge
transfer originating in the Cp ligands, and is in agreement with what
is found for cobaltocene and ferrocene.

On a Cu(111) surface, we find that the molecule binds via dispersion
forces and that the charge transfer is negligible, hence keeping the
above molecular properties.  The molecules present two conformers,
one where the Cp rings are aligned,
eclipsed conformer, and a second conformer where the
Cp are alternatively rotate in a staggered fashion.
We find that systematically the eclipsed conformer is more stable.

The transport properties of the molecules are computed in the tunneling
and contact regimes. On the adsorbed-molecule setup, a second electrode
is approached. We have used an electrode-molecule distance of 5.15~\AA~to
characterize the tunneling regime.  The contact regime corresponds to a
molecule-electrode distance of 2.72~\AA.  We find that the Fermi energy
is in the middle of the HOMO-LUMO gap and that the transmission is largely
dominated by the tail of the LUMO resonance. Due to the large contribution
of the Cp ligands to the majority-spin HOMO and LUMO we find a large
electron transmission for the majority spin channel.  At the same time,
the electron transmission through the minority-spin channel is smaller
due to the prevalence of the TM-d orbitals.  As a result, we find  a
strong spin polarization in the current, with a polarization of  98\%
in the tunneling geometry and 86\% in contact.

When voltage is applied across the molecular junction, we find a
small opening of the HOMO-LUMO gap in the majority-spin channel,
while a negligible effect for the minority-spin one.  The current spin
polarization is very constant, changing from the above 86\% at 0 V to 83\%
at 0.5 V. The behavior with bias is very weak due to the weak coupling
of the molecule to the electrodes and the negligible charge
transfer. However, as the bias increases inelastic channels open that
further reduce the spin polarization of the current.

For biases larger than 1.64 mV, equivalent to the MAE of the molecule, electrons
can flip the molecular magnetic moment out of the easy plane. As a result the
spin of electrons also change and the spin polarization is reduced. For
the first excitation threshold this reduces the current polarization to 33\%.

Also thermoelectric effects in the absence of applied bias lead to a
strong suppression of the otherwise excellent spin-filtering properties
of the molecules. When bias is applied, the much larger
bias contribution overrides the small thermopower and
the spin-filtering properties of the molecular junction are recovered.

{In conclusion, a superficial analysis of our calculations would show  
the triple-decker molecule CoFeCp$_3$ as an excellent current spin filter}. 
However, spin-flip processes and thermocurrents have very negative
consequences for this type of device. A negligible temperature
difference between electrodes can rapidly diminish the spin-filter
efficiency when the electronic transport is governed by thermoelectric
currents. Moreover, ubiquitous spin-flip inelastic effects need to be considered
when evaluating the spin-filtering properties of a molecular junction.

\section*{Acknowledgments}

We acknowledge financial support from Spanish MINECO (Grant
No. MAT2012-38318-C03-02 with joint financing by FEDER Funds from the
European Union). P. A acknowledges finantial support from CONICET. 
ICN2 acknowledges support from the Severo Ochoa Progam (Mineco, Grant SEV-2013-0295)
%\end{acknowledgments}

\begin{suppinfo}

\begin{itemize}
\item Pseudopotential
\item Basis Set
\item Transport calculations for the D5d (staggered) conformer 
\item Lorentzian fits to transmissions and molecular contributions to transport
\item Mulliken analysis as a function of the applied bias
\end{itemize}

\end{suppinfo}

\bibliography{References}

\begin{mcitethebibliography}{52}
\bibitem[Not()]{Note-1}
The clear correspondence between peaks (marked P1, P2, P3 and P4 in
  Fig.~\ref{d2_5_norot} (a)) on the transmission function and the PDOS,
  Fig.~\ref{d2_5_norot}~(b-c), makes it possible to assign P1 at -0.61 eV (P2
  at -0.67 eV) and P3 at 1.18 eV (P4 at -1.42 eV) peaks to the transmission
  trough HOMO$\uparrow$ (HOMO$\downarrow$) and LUMO$\uparrow$
  (LUMO$\downarrow$), respectively. The Fermi level is found to lie closer to
  HOMO states than LUMO states for both spin channels. However, the
  linear-response regime is controlled by the LUMO of both spins. We expect
  then that a low-bias scanning tunneling microscope image will be an image of
  the molecular LUMO.\relax
\mciteBstWouldAddEndPunctfalse
\mciteSetBstMidEndSepPunct{\mcitedefaultmidpunct}
{}{\mcitedefaultseppunct}\relax
\EndOfBibitem
\bibitem[Not()]{Note-2}
The narrower-width and lower-height peak found for P2 as compared to the P1 one
  can be understood by close inspection of the involved molecular orbitals. As
  it was previously discussed, HOMO$\downarrow$ (P2) is basically an $e_2$
  orbital ($d_{xy}-d_{x^2-y^2}$ perpendicular to the transport direction)
  orbital with its electronic cloud completely localized on the TM-centers,
  this leads to a poor hybridization with the electrodes and low transmission
  probability. On the other hand, the HOMO$\uparrow$ (P1) is more delocalized
  and it is built from $e_1$ ($d_{xz}, d_{yz}$) TM-d orbitals. As a result its
  hybridization with the electrodes as well as its transmission probability are
  higher. Concerning peaks P3 and P4, it is interesting to note that although
  both show similar transmission probabilities, the P3 structure has a much
  broader energy range than the P4 one. Such features can be understood taking
  into account that although, LUMO$\uparrow$ (P3) and LUMO$\downarrow$ (P4) are
  both built from TM-d($e_1$) orbitals, LUMO$\downarrow$ is much more localized
  about the TM-centers. As a result, its hybridization with the surface
  electrodes is lower than for the LUMO$\uparrow$.\relax
\mciteBstWouldAddEndPunctfalse
\mciteSetBstMidEndSepPunct{\mcitedefaultmidpunct}
{}{\mcitedefaultseppunct}\relax
\EndOfBibitem
\bibitem[Not()]{Note-3}
This configuration is 0.01 eV more stable than the one with
  $d_1=d_2=2.65$~\AA~where 2.65~\AA~is the average distance of the molecule on
  the surface\relax
\mciteBstWouldAddEndPuncttrue
\mciteSetBstMidEndSepPunct{\mcitedefaultmidpunct}
{\mcitedefaultendpunct}{\mcitedefaultseppunct}\relax
\EndOfBibitem
\bibitem[Not()]{Note-4}
Due to the fact that the eigenvalues of the transmission function corresponding
  to eigenchannels at k$_{xy}=(0,0)$ do not provide the major contribution to
  the total transmission\relax
\mciteBstWouldAddEndPuncttrue
\mciteSetBstMidEndSepPunct{\mcitedefaultmidpunct}
{\mcitedefaultendpunct}{\mcitedefaultseppunct}\relax
\EndOfBibitem
\end{mcitethebibliography}

\end{document}